\documentclass[conference]{IEEEtran}
\IEEEoverridecommandlockouts

\usepackage{amsmath,amsfonts}
\usepackage{amssymb,amsthm}

\usepackage{colortbl}
\usepackage{algorithm}
\usepackage{algorithmicx}

\usepackage{subfigure}

\usepackage[noend]{algpseudocode}
\usepackage{lineno}

\usepackage{array}
\usepackage{textcomp}
\usepackage{url}
\usepackage{verbatim}
\usepackage{graphicx}
\usepackage{cite}

\usepackage{pgfplots}
\pgfplotsset{compat=1.18}
\usepackage{pgfplots}
\usepackage{tikz}

\usepackage{enumerate}
\usepackage{graphicx}
\usepackage{diagbox}

\usepackage{url}
\usepackage{hyperref}
\usepackage{comment}

\usepackage{pifont} 
\usepackage{textcomp}
\usepackage[most]{tcolorbox}
\usepackage{booktabs}

\begin{document}

\title{Invisible Trails? An Identity Alignment Scheme based on Online Tracking}

\author{Ruisheng Shi$^1$,
Zhiyuan Peng$^1$,
Tong Fu$^1$,\\
Lina Lan$^1$,
Qin Wang$^2$,
Jiaqi Zeng$^1$ \\
\textit{$^1$Beijing University of Posts and Telecommunications} $|$
\textit{$^2$UNSW Sydney} \\

}


\maketitle

\begin{abstract}

Many tracking companies collect user data and sell it to data markets and advertisers. While they claim to protect user privacy by anonymizing the data, our research reveals that significant privacy risks persist even with anonymized data. Attackers can exploit this data to identify users' accounts on other websites and perform targeted identity alignment. 
In this paper, we propose an effective identity alignment scheme for accurately identifying targeted users. We develop a data collector to obtain the necessary datasets, an algorithm for identity alignment, and, based on this, construct two types of de-anonymization attacks: the \textit{passive attack}, which analyzes tracker data to align identities, and the \textit{active attack}, which induces users to interact online, leading to higher success rates.
Furthermore, we introduce, for the first time, a novel evaluation framework for online tracking-based identity alignment. We investigate the key factors influencing the effectiveness of identity alignment. Additionally, we provide an independent assessment of our generated dataset and present a fully functional system prototype applied to a cryptocurrency use case.

\end{abstract}

\begin{IEEEkeywords}
Identity Alignment, Online Tracking, De-anonymization, Cookies.
\end{IEEEkeywords}


\section{Introduction}
During the evolution of the internet, third-party tracking technology\vspace{3pt}\footnote{Third-party tracking is conducted by entities that are not the website owners. These trackers are typically added to websites through embedded scripts, ads, or pixels. The collected data is then transmitted to third-party servers. These data tracking companies may integrate the information across multiple websites for purposes such as user analysis. } was introduced to facilitate personalized recommendations, such as targeted advertising \cite{lerner2016internet}. However, online tracking has become ubiquitous, with nearly 90\% of all websites incorporating some form of tracking script, such as cookies. Notably, trackers like Google now have the capability to cover over 80\% of websites \cite{dambra2022sally}.

While these trackers provide convenience, they also raise significant privacy concerns. Users are increasingly worried about data leakage, fearing that their online data might be exploited for commercial or even malicious purposes.

Major companies, such as Google, claim to have implemented advanced anonymization technologies, including generalization and differential privacy, to protect user information and mitigate potential impacts on user privacy \cite{google_anonymization}. Despite these measures, the digital footprints left by users on the internet pose significant threats to their anonymity. Risks such as cookie synchronization, leakage of personal information on social networks, and circumvention of GDPR consent exist, enabling attackers to de-anonymize user identities once they acquire behavioral data \cite{papadopoulos2019cookie}\cite{bekos2023hitchhiker}.

This raises a critical research question:

\smallskip
\textit{Is the ostensibly anonymous personal information from users (e.g., identities and sensitive data) truly safe when collected by online trackers, much like "invisible trails"?}
\smallskip

Unfortunately, our findings indicate that the answer is NO. Even if attackers do not obtain explicit personal attribute information or the content of user-generated activities, they can still perform correlation analysis on user accounts across different websites. This can be achieved by analyzing pseudonymous users' online behavior time records in conjunction with publicly available posting behavior data. To validate our statement, we conducted a series of comprehensive investigations.

\smallskip
\noindent\textbf{Identifying technical challenges (TCs).} Based on extensive investigation, we found a series of technical challenges in the identity alignment process (\textit{trace-and-identify} users) during the data collection and algorithm design stages.

\smallskip
\noindent \textit{\textcolor{red}{$\triangleright$} TC-I: Incomplete datasets for online user behaviors.} Firstly, publicly available datasets are very scarce, with most data being static and single-dimensional. Existing research lacks methods for collecting dynamic user behavior data. Secondly, the timing of web-site event trackers may differ when displayed publicly, causing inaccuracies.

\vspace{-0.3em}
\begin{center}
\tcbset{
        enhanced,
        colback=black!5!white,
        boxrule=0pt,
        colframe=black!75!black,
        fonttitle=\bfseries
       }
       \begin{tcolorbox}
       \let\footnotetext=\footaux
       \small
Evidence 1: only 22.2\% of users use the same username across Facebook and Twitter, with an even smaller percentage using the same username across international platforms. This low overlap demonstrates how user behavior data across platforms is sparse and fragmented, making it hard to build reliable identity mappings.

       \end{tcolorbox}
\end{center}

\smallskip
\noindent \textit{\textcolor{red}{$\triangleright$} TC-II: Inherent diversity in data crawled from different sources.}  The large variations in how different platforms collect and present user attributes result in inconsistent datasets, which complicate cross-platform identity alignment.

\vspace{-0.3em}
\begin{center}
\tcbset{
        enhanced,
        colback=black!5!white,
        boxrule=0pt,
        colframe=black!75!black,
        fonttitle=\bfseries
       }
       \begin{tcolorbox}
       \let\footnotetext=\footaux
       \small
Evidence 2: Different platforms require personal information based on their functionalities. For instance, LinkedIn and Facebook focus more on users' educational backgrounds, while Yelp and Foursquare emphasize users' location information.
       \end{tcolorbox}
\end{center}

\smallskip
\noindent\textcolor{red}{$\triangleright$} \textit{TC-III: Strong assumption for honesty.} State-of-the-art identity alignment methods often assume that users provide honest and valid information. However, when criminals use false information, hide personal details, or tamper with behavior or friend relationships, the effectiveness of these methods is significantly compromised.

\smallskip
\noindent\textcolor{red}{$\triangleright$} \textit{TC-IV: Loose connections in cross-board accounts.} When addressing the cross-border social network account alignment problem, the target node's social network accounts often have low friend association across domestic and international platforms, making it challenging to match identities using friend relationships. Additionally, the target's posting style, content, and purpose on domestic and international social networks can vary significantly, making it difficult to match the identities of criminals based on the similarity of behaviors across platforms.

\smallskip
\noindent\textcolor{red}{$\triangleright$} \textit{TC-V: Poor accuracy.} 
Most existing methods adopt generalized alignment strategies, achieving identity alignment probabilistically, also with low accuracy. Furthermore, these methods primarily rely on passive alignment of user identities, such as collecting user friend information and employing identity alignment algorithms for matching. Consequently, current approaches face challenges in swiftly and accurately identifying user identities.

\smallskip
\noindent\textcolor{red}{$\triangleright$} \textit{TC-VI: Short of evaluation metrics.}
Current research lacks a standardized evaluation framework, with most assessments relying on user profile data scraped from web pages and evaluation metrics primarily derived from machine learning. This results in inconsistent evaluations and difficulties in transferability, ultimately failing to provide accurate references for different communities.


\smallskip
\noindent\textbf{Our new design.} To overcome aforementioned challenges, we design and propose an innovate identity alignment scheme for accurately identify targeted identity. We highlight our design spotlights (i.e., \textit{\textbf{key modules}}).

\smallskip
\noindent \textcolor{teal}{$\triangleright$ \textit{A data collector} } (Sec.\ref{sec-data}, addressing \textit{TC-I\&TC-II}). We present an effective data collector capable of covering a wide range of data categories, including anonymous, incomplete, and cross-domain data. This collector can collect publicly available web data (via crawlers) and also anonymous behavioral data (via trackers). In scenarios where the target user data is minimal, we can actively interact with users online, encouraging them to frequently post to increase the data volume. Our core data generation algorithm (cf. Algorithm~\ref{alg1:generate}) not only facilitates the characterization of the collected data but also ensures high-quality data output. Our approach remains effective and accurate even when the target user information is sparse or when there are significant structural differences between the target and the source websites.

\underline{\textbf{\textit{Add-on 1:}}} To validate the accuracy of our generated datasets and deepen our understanding, we have conducted independent evaluations and included these specialized assessments as an additional section (Sec.\ref{sec-datasetEva}).

\smallskip
\noindent \textcolor{teal}{$\triangleright$  \textit{An identity alignment scheme} } (Sec.\ref{sec-align}, addressing \textit{TC-III\&TC-IV\&TC-V}). We propose a novel algorithm for identity alignment (Algorithm~\ref{alg2:Identity}). Building on a dataset collected from multiple sources, our algorithm achieves high-precision identification of specific anonymous user addresses, maximizing the matching set (Algorithm~\ref{alg3:Matching}). Additionally, our algorithm supports diverse scenarios, including the alignment of single or multiple accounts, cross-border account alignment, cross-device alignment, and the integration of both open and anonymous data.

\smallskip
\noindent \textcolor{teal}{$\triangleright$ \textit{Two attacking methods} } (Sec.\ref{sec-attack}). We accordingly propose two attacking methods: (i) \textit{A passive attack} (Sec.\ref{sec-passiveattack}) outputs a minimal set of identities that meets the requirement. This attack leverages our identity alignment scheme to identify target identities by comparing and analyzing user data across different websites, encompassing both anonymous user behaviors and public information collected from social networks. (ii) \textit{An active attack} (Sec.\ref{sec-activeattack}) builds on our passive attack. In addition to the methods used in passive attacks, active attacks study the interests and hobbies of targeted users (e.g., cyber criminals), actively inducing them to interact online to gather more information and ultimately identify the target identity.

\smallskip
\noindent\textbf{New evaluation framework} (Sec.\ref{sec-metric}, addressing \textit{TC-VI}). We introduce a novel evaluation framework specifically designed for identity alignment. We propose three customized evaluation metrics for identity alignment algorithms (Sec.\ref{sec-algnmetri}) that more accurately reflect their success rates and accuracy in typical scenarios. Additionally, our evaluation of the tracker datasets involves extracting a range of features (Sec.\ref{sec-datasetEvaResults}, e.g., the ratio of browsing to posting, time offsets in browsing, the distribution of page views, and time differences in posting records), all of which contribute to improving accuracy.

We perform experiments (Sec.\ref{sec-experiment}) and present a series of key home-taking findings (\textbf{Result}-\ding{202}-\ding{204}).

\smallskip
\noindent\textbf{Practical applications.} We further introduce two practical applications where our identity alignment scheme can be effectively applied.

\smallskip
\noindent $\triangleright$ \textit{Crypto-based criminals.}  Cryptocurrencies are a form of distributed digital currency with strong anonymity features. The use of cryptocurrencies for illegal transactions and criminal activities, such as cross-border money laundering, drug trafficking, and cyber fraud, has been increasing. Law enforcement agencies find it challenging to effectively trace these activities, primarily because the real identities of the criminals cannot be accurately identified. Our system can identify and track cryptocurrency criminals through features such as identity alignment, suspicious address verification, and the association of identity information with cryptocurrency addresses, thereby effectively linking virtual addresses to physical locations.

To elaborate on the technical workflow, our scheme achieves criminal tracing through multi-layered association mechanisms: Specifically, our scheme enables the tracing of criminals' identities by linking suspicious cryptocurrency addresses to their real identities, thereby mitigating the risk of malicious exploitation of cryptocurrency anonymity. For instance, given a suspicious cryptocurrency address and its associated overseas accounts, the identity alignment algorithm can match the suspect's real-name domestic account to obtain their authentic identity information. Meanwhile, the data collector of our system crawls location-related information associated with the domestic account, such as IP address attribution and mobile phone registration address, which further connects the virtual cryptocurrency address to the criminal's physical location, achieving location tracing of the criminal.

\underline{\textbf{\textit{Add-on 2:}}} To put our scheme into practice, we have developed a prototype system for identifying crypto suspects involved in criminal activities (Sec.\ref{sec-cryptoPrototype}) and demonstrated the system’s efficacy and feasibility.

\smallskip
\noindent $\triangleright$ \textit{Tor-based anonymination.} Tor (the Onion router) facilitates anonymous communication by routing internet traffic through a series of nodes, called relays. Each relay decrypts a layer of encryption to reveal the next relay in the path, preventing any single point from knowing both the sender's and recipient's locations. While Tor is often used to access the ``dark web'' for legitimate purposes like protecting privacy or evading censorship, our scheme aids in deanonymizing the user's location and usage, countering network surveillance or traffic analysis.

\section{Technical Warm Ups}
\label{sec-system}

\noindent\textbf{Identity alignment.}
Digital identity refers to the creation of an online persona or identity in the digital world~\cite{bentley2020social}. The research process of identity alignment technology is mainly divided into two stages: social network user \textit{data collection} and the design of \textit{identity alignment algorithms} for different users. Specifically, the first stage involves collecting real user data from social networks. This includes, but is not limited to, usernames, profile pictures, posted content, and the relationships between platform users to obtain the unique ``imprint" of users on social networks. The second stage, the identity alignment algorithm, is the core aspect of identity alignment research. Based on the data collected in the first stage, this phase utilizes algorithmic models to predict and match user relationships across different social networks. These algorithms determine whether different social network imprints point to the same real-world entity by comparing the similarity of user imprints across various social networks.

\smallskip
\noindent\textbf{Web tracker.}
Web tracking can be implemented in various ways, including online tracking methods such as cookies, browser fingerprinting, and flash local shared objects.

Cookies were created to identify returning users and maintain online shopping carts\cite{MDN}. Cookie tracking is a method where, when a user opens a browser and visits a website for the first time, the website provides a cookie to record the user's visit. This cookie is stored on the user's device by their web browser. When the user opens the site again, the cookie information is sent to the site along with the access request, allowing the site to recognize the user and record their visit.

However, third parties soon began using cookies to track users across websites and serve targeted ads\cite{DoubleClick}. The early standardization efforts focused on preventing unintended cookie sharing across domains but largely ignored the intentional misuse of cookies by third parties for cross-site tracking\cite{Smartcookie}. As a result, the use of third-party cookies for cross-site tracking has become widespread\cite{aqeel2020landing}\cite{chen2018mystique}\cite{pochat2018tranco}\cite{randall2022trackers}. Studies have found that the majority of third-party cookies are set by advertising\cite{chen2018mystique} and tracking services and that third-party cookies outnumber first-party cookies by a factor of two\cite{aqeel2020landing}, or even up to four when they contain identifiers\cite{randall2022trackers}. Advertising networks and data proxies use a range of techniques to track users on the web, including standard stateful cookie-based tracking and stateless fingerprinting.

\section{System Model}
\label{sec-system}

\subsection{Roles, Threat Model, And Goals}

\noindent\textbf{Roles.} We have identified five entities.

\begin{itemize}
\item \textit{Target user:} An entity registered and published posts on both the source and target websites.
\item \textit{SiteA (source website):} A platform that supports user posts. It is established that the attacker is aware of the target user's account on this website.
\item \textit{SiteB (target website):} Similar to SiteA, this website supports user posts. The target user's account on SiteB is unknown to the attacker.
\item \textit{SiteT (tracking website):} Trackers collect user web behavior data through online tracking.
\item \textit{Attacker:} An individual with knowledge of the user's account on SiteA, has obtained the tracking dataset collected by tracker.
\end{itemize}

\noindent\textbf{Goal.} Assuming that the target user's account on the source website has been obtained, and that the tracker website can record the user's online behavior on both the source and target websites through a cookie synchronization mechanism, our identity alignment algorithm is expected to be able to deanonymize such a user: i.e., it is possible to identify the target user's account on the target website based on analyzing the information posted by the target user on the source website and the anonymous user behavior data provided by the tracker.

\begin{figure}[!hbt]
  \centering
    \vskip -0.1in
  \includegraphics[totalheight=2.3
  in]{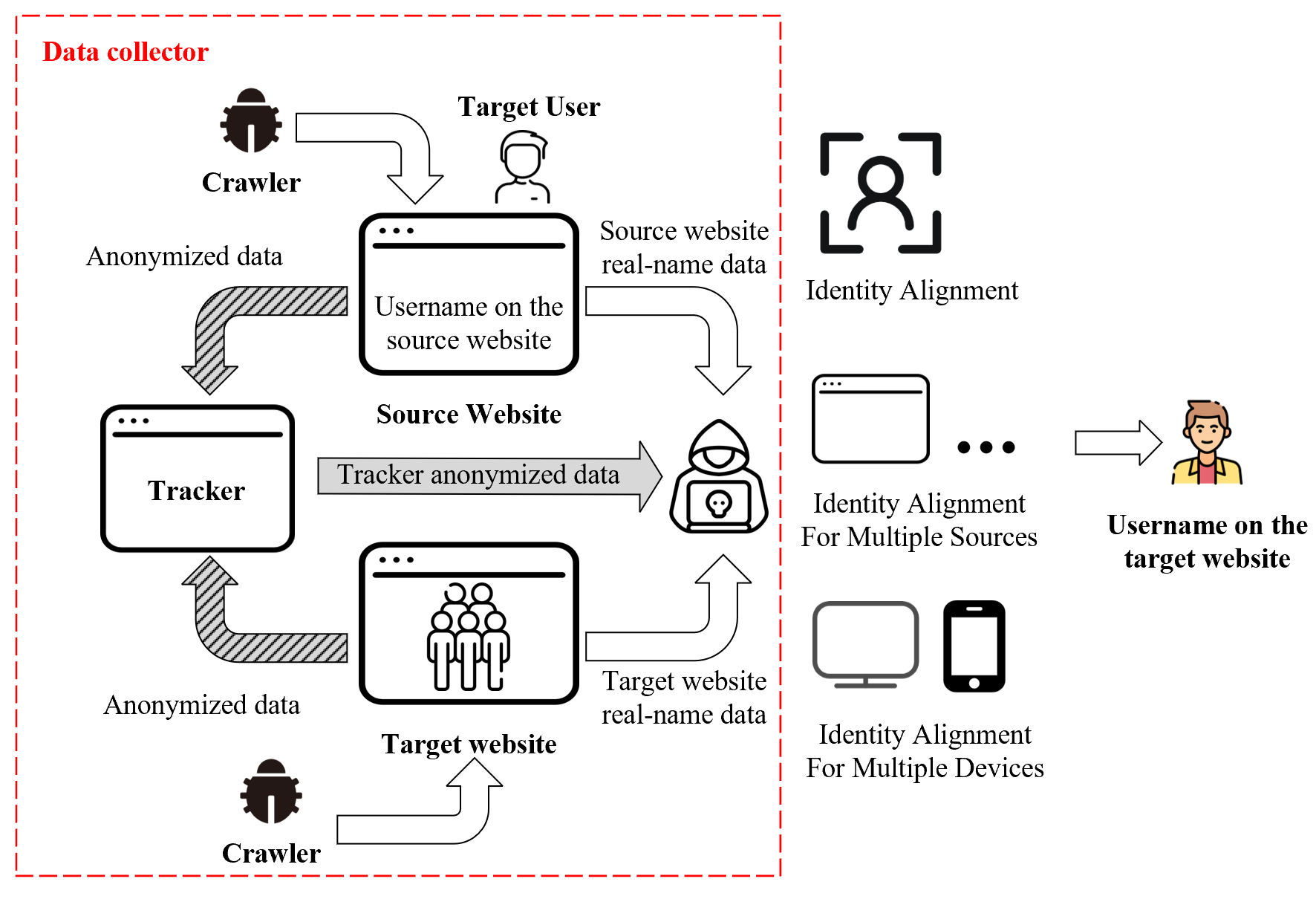}
    \vspace{-0.15in} 
  \caption{Attack Scenarios} \label{fig:Senario}
  \vskip -0.15in
\end{figure}

\smallskip
\noindent\textbf{An instance} (Fig.\ref{fig:Senario}). We assume there is a criminal suspect (target user) who may be active on both an anonymous foreign data website (SiteA, source website) and a real-name domestic website (SiteB, target website). 

SiteA hosts a large number of user accounts and their information, including posts, likes, and shares. Since SiteA is not governed by domestic law enforcement, authorities cannot directly obtain the real identities of its users. Meanwhile, a tracker collects user behavior data from both foreign and domestic sites and anonymizes it. SiteB, unlike the foreign site, operates under domestic cybersecurity laws, which often require real-name verification. Thus, if the suspect's account on SiteB can be identified, their real identity within the country can be traced.

Assuming the suspect’s account on the foreign site has been identified, and the tracker can synchronize user behavior data across SiteA and SiteB through cookie mechanisms. In this scenario, a tracker-based identity alignment algorithm (SiteT, tracking website) can de-anonymize the target user by analyzing the information posted on SiteA and the anonymized behavior data provided by the tracker. This can reveal the target user's account on SiteB and then accurately identify the suspect.

\smallskip
\noindent\textbf{Threat model.} We assume two types of adversaries. (i)
\textit{passive adversaries:} We consider passive adversaries to be those who can only crawl, collect, track, and analyze online data. These adversaries launch passive attacks by using the collected data and applying specific matching algorithms. (ii) \textit{active adversaries:} They are capable of launching targeted inducements. For example, they may post relevant information that interests targeted users on social websites to lure them into frequently posting or interacting with the content.

Passive adversaries do not engage in direct interference with users' online activities. Instead, they rely on collecting and analyzing user behavior to narrow down the pool of users to a smaller group of suspicious individuals for further scrutiny. In contrast, active adversaries can enhance the precision of this process. They do so by employing targeted strategies to refine the suspect pool, thereby making it smaller and more accurate, which facilitates the efficient and precise identification of the targeted user.

\subsection{Key Modules}

\noindent \textbf{Data collector.} Our data collector consists of two components: data crawler and data tracker. The module can crawl, collect and handle the public data from different website (via a crawler) and make them into the desired datasets (via trackers) for the subsequent algorithms.  

We can collect two types of data: (i) \textit{anonymity dataset} (behavioral trails without identities), (ii) \textit{public dataset} from SiteA (the source website) and from SiteB (the target website) with user's identities. 

The anonymity dataset primarily includes timestamps and corresponding domain names for all activities, such as publication dynamics and web browsing of users from both the target site and the source site. Pseudonyms are used as tracking identifiers for the same entity. These tracking identifiers are independent of website usernames and real identity information. Consequently, while this data can record users' online behavioral trajectories, it is challenging to correlate it with real-world identities or the behavioral information of users on the source and target sites.

The public dataset from SiteA includes usernames and timestamps of dynamic posts published on the source website.  SiteB data includes timestamps of all users' dynamic behaviors published on the target website, offering a detailed view of user interactions on SiteB.

\smallskip
\noindent \textbf{Identity alignment algorithm.} Our algorithm identifies potential usernames on a target website by combining publicly available webpage data with tracker data. It starts by collecting the times of a target user's activities on a source website, filtering these within a specific time window. These times are then matched with tracker data to identify tracking IDs that correspond to the user's activities. The algorithm uses these IDs to gather times of activity on the target website and attempts to match them with the activity times of other users on that site. If a match is found within a defined time-matching range, the corresponding usernames are considered potential matches, effectively aligning identities across different platforms (either domestically or abroad). 

\smallskip
\noindent\textbf{Design spotlights.} We highlight our several benefits.

\begin{itemize}
    \item \textit{High accuracy and effectiveness}: Our design can successfully deanonymization the target user based on online tracking. Even trackers often utilize irreversible encryption and pseudonymization (i.e., GDPR standards), complicating user identification within datasets, anonymized datasets still contain sufficient deanonymization information (e.g., aligning usernames between target and source websites) to link online actions across different platforms.

    \item \textit{Multi-scenarios}: We have designed our algorithm to optimize the time-matching parameter. Our experiments account for various scenarios involving time discrepancies, including those where user actions are recognized and those where they are not, ensuring a comprehensive evaluation.

    \item \textit{Multi-device usage}: Aligning user identities across multiple devices presents significant challenges. For example, Taobao and iQiyi use cookies or URL parameters to track users across devices. Our method is designed to address these challenges and facilitate effective cross-device identity alignment.
\end{itemize}

\section{Making Datasets}
\label{sec-data}

In this section, we build datasets by collecting online data across multiple websites. Our collection covers two datasets, a public dataset based on specified crawling technique and a tracker dataset that majorly captures behavioral data. 



\subsection{Building Public Dataset}

We begin by detailing our data crawling process for public datasets (short for \textbf{P}). We use automated web crawler through Selenium automation tools to collect user posting behavior data on real social networks, mainly by randomly selecting user names through keywords and then crawling the home pages of these users. The processes are as follows:

\begin{itemize}
\item[P1] \textit{Keyword search.} Build a list of user ids by automatically crawling a specific keyword and randomly selecting users mentioning the keyword in a post.

\item[P2] \textit{User page crawling.} The user ID collected in the previous step is concatenated with "twitter.com" to obtain "twitter.com/<username>". Crawl in depth for each user, focusing on collecting timestamps of tweets posted on the user's page.

\item[P3] \textit{Simulated scrolling.} To ensure that we could collect more data about the user's Posting behavior on the homepage, We automatically simulated scrolling. Specifically, the automatic crawling is used to imitate the action of the artificial mouse down East China, and the scrolling is performed every 5 seconds to trigger the loading of the user's homepage post content, which can deepen the depth of the user's homepage crawling information.

\item[P5] \textit{Page stay.} Since continuous page requests may trigger the anti-crawler mechanism, we set it to stay on each user's home page for 30 seconds, on the one hand, to simulate the browsing behavior of real users, and on the other hand, to avoid the risk of being detected as an automated crawler.
\end{itemize}

\subsection{Building Anonymity Dataset}

We then build the tracker's anonymity dataset (short for \textbf{T}) with a core tracking dataset generation algorithm.

\smallskip
\noindent\textbf{Parameter formation.}
Based on our investigation, we first outline the key parameters that make up the tracker dataset. We focus on four critical parameters, assuming that the tracker collects only minimal user information.

\begin{itemize}
\item \textit{Distribution of number of views.}
We use the Pareto distribution to represent the distribution of the number of views of each user's posts because it reflects how most events are concentrated in a small number of groups. We also add random numbers to account for natural browsing behavior.

\item \textit{Distribution of browsing time offset.}
The power-law distribution describes how the probability of certain events, like posts being viewed, decreases as their value increases. This distribution can simulate more realistic user browsing behavior patterns.

\item \textit{Browse the release multiples $n$.}
In reality, user browsing behavior is far more than dynamic publishing behavior, and the ratio of tracker browsing behavior to publishing dynamic behavior is called browsing publishing multiple.

\item \textit{Time difference $\Delta t$.}
Where the time when the user posted the activity is $t1$, the time when the user behavior was recorded by the tracker is $t2$, and the time difference between the time on the web page and the time recorded by the tracker is  $\Delta t$.
\end{itemize}

\smallskip
\noindent\textbf{Tracker data generation algorithm.}
Next, we describe our algorithm for generating tracker data (cf. Algorithm~\ref{alg1:generate}). We assume a browse-release ratio \( n \) and a time difference \( \Delta t \). The number of views is modeled using a normal distribution with a mean of \( n \) and a standard deviation of 1. Additionally, the browse offset follows a power-law distribution. 

Our algorithm consists of two major processes.

\begin{itemize}
    \item[T1] \textit{Generate tracking IDs:} Establish a one-to-one correspondence between users on {SiteA} and {SiteB} by generating unique tracking IDs. For each user, create dynamic posting behavior data by adding behavior times from {SiteA} and {SiteB} with random time offsets to the tracker dataset.
    
    \item[T2] \textit{Generate tracker's anonymity data:} For each user, create a number of browsing activities based on a normal distribution. Add a small amount of randomness to these counts for each post to reflect natural variability in user behavior. Then, create a set of time intervals, or offsets, using a power-law distribution. The number of these offsets should match the number of dynamic posts made by the user. These time offsets represent the intervals between when users view different posts. Next, Apply the generated time offsets to determine when users view each post. Assign a random tracking ID to each user who views a post to simulate the browsing behavior. Finally, integrate the simulated browsing times and the dynamic posting behavior data to form the complete tracker dataset.

\end{itemize}

In the T1 phase, it is necessary to generate unique tracking IDs for each user ($N$ users) and add random time offsets to each of their posting behaviors ($M$ posts). This process involves traversing all users' posting times, with a complexity of $O(N + M)$. In the T2 phase, it is required to add simulated browsing counts for each posting behavior of each user. The traversal scale at this stage is the product of the total number of user posting behaviors and the browsing-to-publish ratio \((n)\), i.e., \(O(n×(N + M))\). The space complexity of this algorithm is mainly determined by the storage size of temporary lists such as dataTracking, tracking ID, and BrowseCounts, which is proportional to the source website dataset, target website dataset, and browsing-to-publish ratio n. Therefore, the space complexity is \(O(n×(N + M))\).

\begin{algorithm} [!htb] 

\caption{Tracker data generation algorithm} 
\label{alg1:generate}  

\renewcommand{\arraystretch}{1.30}
\footnotesize 
\setlength{\tabcolsep}{4pt} 

\begin{algorithmic}[1]
\Require n \Comment{\textcolor{teal}{browsing-to-publish ratio}}\\
$\Delta t$ \Comment{\textcolor{teal}{The difference between when a post is posted on a web page and when a tracker records it}}\\
$dataA,dataB$: ($A_c, t_1, t_3, \ldots$), e($B_c, t_1, t_2, t_2, \ldots$)\\ \Comment{\textcolor{teal}{user's posting behavior time on the target website}} 

\Ensure $\text{dataTrackingExp}$ \Comment{\textcolor{teal}{Newly generated tracker data}}
 
\Statex \rule{\linewidth}{0.4pt}
\State $dataTracking \gets []$ \Comment{\textcolor{teal}{consists of time of all behaviors on websites}}
\State $current\_time \gets now$
\State $trackingIDs \gets$ hash set with max(len(dataA.users), len(dataB.users))
\State $a, b, i \gets 0$
\ForAll{usera in $dataA$}
    \ForAll{time in $dataA[usera]$}
        \State add $time - random(\Delta t)$ to $dataTracking[trackingIDs[a]]$
    \EndFor
    \State $b \gets b + 1$
\EndFor 
\State $BrowseCounts \gets$ Perato distributed set of $n \times$ length($dataA + dataB$) \Comment{\textcolor{teal}{browsing behavior recorded by the tracker}}
\State $dataTrackingExp \gets []$
\While{$i <$ len($dataTracking$)}
    \State $j \gets 0$
    \While{$j <$ $BrowseCounts[i]$}
        \State $random\_user \gets$ random($trackingIDs$)
        \State $time\_offset \gets$ ($current\_time - dataTracking[i].time$)
        \State $new\_time \gets dataTracking[i].time + time\_offset$    
        \State add $new\_time$ to $dataTrackingExp[random\_user]$
        \State $j \gets j + 1$
    \EndWhile
    \State $i \gets i + 1$
    \State $dataTrackingExp \gets dataTrackingExp + dataTracking$
\EndWhile
\end{algorithmic} 
\end{algorithm}

\section{Our Identity Alignment Scheme}
\label{sec-align}

In this section, we present our identity alignment scheme based analysing the collected dataset with the aim to positioning the targeted users.

\subsection{Parameter Selection}

The identity alignment scheme leverages behavioral data captured by trackers, which record user activities. Critical parameters for consideration include the starting time and duration of the observation window for user behavior, as well as the granularity of time matching\footnote{Our dataset evaluation (Sec.\ref{sec-datasetEva}) reveals discrepancies between the timestamps of user posts as displayed on web pages and those recorded by trackers, necessitating the adoption of time-matching granularity.}.
Consequently, our scheme primarily focuses on the following key parameters:

\begin{itemize}
    \item \textit{Time window length} and \textit{starting time}: The time window length refers to a period of time, with the starting time determining the beginning boundary of the time window. Within the given time window range starting from the starting time, the time set for searching is determined. For example, with a starting time of Jan. 1st, 2024, and a time window of 1 day, we select the set of user post times on Jan. 1st, 2024, as the query condition. This time window can be adjusted according to user behavior, such as selecting a day with high user activity as the query condition.
    
    \item \textit{Time matching granularity}: Time within the defined time-matching granularity is treated as a matching range. In other words, if the timestamp of a behavior on a webpage and the corresponding time recorded by the tracker fall within the same time-matching granularity, they are considered a match. For instance, if the time-matching granularity is set to one minute, then a post on the webpage and the tracker's record will be considered matched if they occur within the same minute.
\end{itemize}

The length of the time window, starting time, and time-matching granularity are critical parameters in data analysis. These factors enable the filtering and pinpointing of data within specific timeframes, thereby enhancing the accuracy and utility of the analysis. In practice, these parameters should be adjusted flexibly according to different scenarios and user conditions to better meet the requirements of analysis and matching, ultimately improving the identification of potential target usernames on the target website.

\subsection{Identity Alignment}

We present the core identity alignment processes in Fig.\ref{fig:attacktwo} (short for \textbf{I}). It involves preprocessing the data to filter timestamps relevant to specific usernames, combining tracking data with time granularity and narrowing down behavior times corresponding to specific usernames on the target website. The final result achieves the user-matching objective from the source to the target websites.

The key processes are outlined in Algorithm~\ref{alg2:Identity}.

\begin{itemize}
    \item[I1] \textit{Search for matching timestamps}: Based on the target website username and query time conditions, including start time, time window, and time granularity, filter out corresponding user behavior timestamp lists (\textit{tA\_lists}) from the source website.
    
    \item[I2] \textit{Find matching tracking IDs in the anonymity dataset}: Check whether each user's behavioral data on the source website in the anonymity dataset fully includes the query condition time with added time matching granularity. If matched, mark them as candidate users and add their Tracking IDs to the candidate Tracking ID set. Remove users with only partial matches from the set, resulting in a list of user Tracking IDs (\textit{id\_lists}). Use these Tracking IDs to find their behavior timestamp lists (\textit{tB\_lists}) in the tracker's anonymity data.
    
    \item[I3] \textit{Search for candidate usernames on the target website.} By comparing these timestamp lists with user behavior timestamps on the target website, identify users with complete matches and adds their usernames to the candidate list.
    \item[I4] \textit{Return.} The returned list of usernames to be investigated contains all matched candidate usernames.
\end{itemize}

In our algorithm, the data is preprocessed to establish matching criteria by selecting a list of behavior timestamps ($\textit{tA\_lists}$) within a specific time window under the target user's username from the source website data ($\textit{dataA}$). The algorithm then identifies online tracking identifiers ($\textit{TID}$) from the tracking data ($\textit{dataTracking}$) that match these timestamps, creating a list of candidate anonymous users ($\textit{id\_lists}$). Subsequently, it uses these identifiers to gather behavior timestamps from the target website ($\textit{dataB}$). By comparing these timestamps with the behavior times recorded for the candidate anonymous users in the tracker data ($\textit{tB\_lists}$), the algorithm identifies usernames on the target website whose posting times completely match. This final list of matched usernames is used to align identities from the source website to the target website, effectively identifying potential usernames across platforms.

Here, the primary factor affecting the time complexity is the traversal of the source website data and the tracker (anonymity) dataset. Therefore, the time complexity is proportional to the size of these two datasets. Additionally, the number of posting behavior timestamps in the source website data that match the given target user also impacts the complexity. 
The space complexity of the algorithm is mainly determined by the storage requirements for temporary results such as \(\textit{tA\_lists}\), \(\textit{id\_list}\), and \(\textit{tB\_lists}\). Since the size of these lists is directly related to the size of the input data, the space complexity is proportional to the amount of input data.
We utilize hash tables to store the source website data and tracking dataset, facilitating the efficient lookup of matching tracking IDs and their storage in \(\textit{id\_list}\). As a result, the complexity of the algorithm approaches \(O(n)\).


\subsection{Identity Alignment For Multiple Sources}

The algorithm for aligning identities across multiple sources (cf. Algorithm \ref{alg3:Matching}) involves identifying a target user through different usernames (\textit{Ac} and \textit{Dc}) on two distinct source websites (\textit{siteA} and \textit{siteD}). 

We match the tracking identifiers associated with \textit{Ac} and \textit{Dc}, denoted as \textit{id\_listA} and \textit{id\_listD}, respectively, from the network user behavior dataset of the tracker. By intersecting these two lists, we obtain a consolidated TrackingID list (\textit{id\_list}) that the identity alignment algorithm uses to match user behavior data on the target website (\textit{siteB}). The specific steps are outlined as follows:

\begin{itemize}
    \item \textit{Input}: Usernames of the target user on two source websites, \textit{UserName1} and \textit{UserName2} from \textit{siteA} and \textit{siteD}, respectively, along with their respective query start times (\textit{t1} and \textit{t2}) and query time windows (\textit{t\_window}). Additionally, three datasets (\textit{dataA}, \textit{dataB}, and \textit{dataD}) containing timestamps of user postings on the source websites are provided.
    
    \item \textit{Extraction of Query Conditions}: The algorithm extracts user activity timestamps within the query time window for each source website’s data, resulting in time lists \(\mathcal{L}_{A}\) and \textit{tD\_lists}.
    
    \item \textit{Identity Alignment}: Using the time lists (\(\mathcal{T}_{A}\) and \textit{tD\_lists}), the algorithm matches the tracking ID lists (\(\mathcal{I}_{A}\) and \textit{id\_listD}) for the target user on \textit{siteA} and \textit{siteD} in the tracker data.
    
    \item \textit{Intersection Matching}: The algorithm compares the two tracking ID lists (\(\mathcal{L}_{id}\)), taking their intersection to identify IDs present in both lists, which suggests potential ownership by the same user.
    
    \item \textit{Generation of Inspection List}: Finally, using the identity alignment algorithm based on network tracking data, the algorithm filters users from \textit{siteB} based on the obtained ID list to generate the final list of candidate usernames for inspection.
\end{itemize}

The time complexity of Algorithm 3 is primarily influenced by the traversal of the source website dataset and multiple calls to Algorithm 2. Specifically, the timestamp extraction operation requires a linear scan of the data record set (with a size of $n$), and the average time complexity of each call to Algorithm 2 is \(O(n)\). Thus, the overall time complexity remains \(O(n)\).In terms of space complexity, it is mainly determined by the temporary storage of time lists (\(\mathcal{T}_{A}\) and \textit{tD\_lists}) and intermediate tracking ID lists (\(\mathcal{I}_{A}\) and \textit{id\_listD}). Dominated by the scale of the dataset, it can be simplified to a linear space complexity of \(O(n)\).

\begin{algorithm}[t]
\caption{Identity Alignment}
\label{alg2:Identity}

\renewcommand{\arraystretch}{1.30}
\scriptsize 
\setlength{\tabcolsep}{4pt} 

\begin{algorithmic}[1]
\Require UserName \Comment{\textcolor{teal}{username of target user in source website}} \\
$t_0$  \Comment{\textcolor{teal}{searching start time}} \\
$t_{\text{window}}$ \Comment{\textcolor{teal}{Size of the query time window (in days)}} \\
$\Delta t_m$ \Comment{\textcolor{teal}{time matching range (in seconds)}} \\
$dataA, dataB$: ($A_c, t_1, t_2, t_3, \ldots$), ($B_c, t_1, t_2, t_3, \ldots$)  \\ \Comment{\textcolor{teal}{user's posting behavior time on the target website}} \\
$dataTracking$: $\{TID; A, t_1, t_2, t_3, t_4, \ldots; B, t_1, t_2, t_3, t_4, \ldots\}$  \\ \Comment{\textcolor{teal}{consists of time of all behaviors on websites}}

\Ensure $\text{MatchedUsers}$ \Comment{\textcolor{teal}{candidate target website usernames}}

\Statex \rule{\linewidth}{0.4pt}

\Function{CollectTimes}{UserName, $t_0, t_{\text{window}}, dataA$}
    \State $tA\_lists \leftarrow []$ \Comment{\textcolor{teal}{Time list to be filled for A}}
    \ForAll{data in $dataA$}
        \If{data.username == UserName}
            \ForAll{time $t_A$ in data.time}
                \If{$t_A$ in [$t_0 - t_{\text{window}}, t_0 + t_{\text{window}}$]}
                    \State add $t_A$ to $tA\_lists$
                \EndIf
            \EndFor
        \EndIf
    \EndFor
    \State \Return $tA\_lists$
\EndFunction

\Statex \rule{0.9\linewidth}{0.4pt}

\Function{MatchTrackingIDs}{$tA\_lists, \Delta t_m, dataTracking$}
    \State $id\_lists \leftarrow []$ \Comment{\textcolor{teal}{ID list to be filled for A}}
    \ForAll{data in $dataTracking$}
        \If{data.domain == 'siteA'}
            \ForAll{time in data.times}
                \ForAll{$t_{\text{target}}$ in $tA\_lists$}
                    \If{time in [$t_{\text{target}} - \Delta t_m$, $t_{\text{target}} + \Delta t_m$]}
                        \State add data.trackingID to $id\_lists$
                    \EndIf
                \EndFor
            \EndFor
        \EndIf
    \EndFor
    \State \Return $id\_lists$
\EndFunction

\Statex \rule{0.9\linewidth}{0.4pt}

\Function{CollectTimesFromIDs}{$id\_lists, dataTracking$}
    \State $tB\_lists \leftarrow []$
    \ForAll{data in $dataTracking$}
        \If{data.trackingID in $id\_lists$ and data.domain == 'siteB'}
            \State add data.times to $tB\_lists$
        \EndIf
    \EndFor
    \State \Return $tB\_lists$
\EndFunction

\Statex \rule{0.9\linewidth}{0.4pt}

\Function{MatchUsers}{$tB\_lists, dataB, \Delta t_m$}
    \State $MatchedUsers \leftarrow []$ \Comment{\textcolor{teal}{List to be filled for matched users}}
    \ForAll{times in $tB\_lists$}
        \ForAll{data\_item in $dataB$}
            \State $\text{match} \leftarrow \text{true}$
            \ForAll{time $t$ in data\_item.time}
                \If{$t \not\in [t - \Delta t_m, t + \Delta t_m]$}
                    \State $\text{match} \leftarrow \text{false}$
                    \State \textbf{break}
                \EndIf
            \EndFor
            \If{$\text{match}$}
                \State add data\_item.username to $MatchedUsers$
            \EndIf
        \EndFor
    \EndFor
    \State \Return $MatchedUsers$
\EndFunction
\end{algorithmic}
\end{algorithm}

\begin{algorithm}[t]
\caption{Matching Algorithm for Multiple Accounts}
\label{alg3:Matching}

\renewcommand{\arraystretch}{1.30}
\scriptsize 
\setlength{\tabcolsep}{4pt} 

\begin{algorithmic}[1]
\Require $UserName1$ \Comment{\textcolor{teal}{Username for siteA}}
\Require $UserName2$ \Comment{\textcolor{teal}{Username for siteD}}
\Require $t1, t2$ \Comment{\textcolor{teal}{Start time for accounts on siteA and siteD}}
\Require $tw$ \Comment{\textcolor{teal}{Size of the query time window (in days)}}
\Require $dataA$: $(A_c, t_1, t_2, t_3, \ldots)$ \Comment{\textcolor{teal}{Data from the source website siteA}}
\Require $dataB$: $(B_c, t_1, t_2, t_3, \ldots)$ \Comment{\textcolor{teal}{Data from the target website siteB}}
\Require $dataD$: $(D_c, t_1, t_2, t_3, \ldots)$ \Comment{\textcolor{teal}{Data from the source website siteD}}

\Ensure $\text{matched\_users}$ \Comment{\textcolor{teal}{Candidate usernames}}

\Statex \rule{\linewidth}{0.4pt}

\State $\mathcal{T}_{A} \leftarrow []$ \Comment{\textcolor{teal}{Time list to be filled for A}}
\State $tD\_lists \leftarrow []$ \Comment{\textcolor{teal}{Time list to be filled for A}}

\ForAll{data in $dataA$} \Comment{\textcolor{teal}{Extract query conditions in siteA}}
    \If{data.username == $UserName1$}
        \ForAll{time $tA$ in data.time}
            \If{$tA$ \textbf{within} $[t1 - tw, t1 + tw]$}
                \State add $tA$ to $\mathcal{T}_{A}$
            \EndIf
        \EndFor
    \EndIf
\EndFor

\ForAll{data in $dataD$}
    \If{data.username == $UserName2$}
        \ForAll{time $tD$ in data.time}
            \If{$tD$ \textbf{within} $[t2 - tw, t2 + tw]$}
                \State add $tD$ to $tD\_lists$
            \EndIf
        \EndFor
    \EndIf
\EndFor

\State $\mathcal{L}_{A} \leftarrow Algorithm\ 2(\mathcal{T}_{A})$  \Comment{\textcolor{teal}{Obtain the account's trackingID lists in tracker}}
\State $id\_listD \leftarrow Algorithm\ 2(tD\_lists)$

\ForAll{id in $\mathcal{L}_{A}$} \Comment{\textcolor{teal}{Obtain the intersection of two trackingID lists}}
    \If{id in $id\_listD$}
        \State add id to $\mathcal{L}_{id}$
    \EndIf
\EndFor

\State $matched\_users \leftarrow Algorithm\ 2(\mathcal{L}_{id})$

\end{algorithmic}
\end{algorithm}

\subsection{Identity Alignment For Multiple Devices}

In practice, users often access services across multiple devices. Websites that implement deterministic cross-device tracking technology can recognize when a user logs in with the same credentials on both a mobile device and a PC. In such cases, the website assigns a shared identifier to both devices. This shared identifier enables the tracker to aggregate and analyze the user’s behavior data across devices, using methods such as \textit{cookies} or \textit{URL} parameters to maintain continuity. For example, Taobao and iQiyi employ deterministic cross-device tracking. Taobao's tracker, \textit{mmstat}, utilizes \textit{cookies} to link user identities across different devices, while iQiyi uses \textit{URL} for the same purpose.

\section{Two Attacks}
\label{sec-attack}
Based on our identity alignment algorithm, we can establish two types of attacks. 

\begin{figure*}[!hbt]
  \centering
  \includegraphics[width=0.9\textwidth]{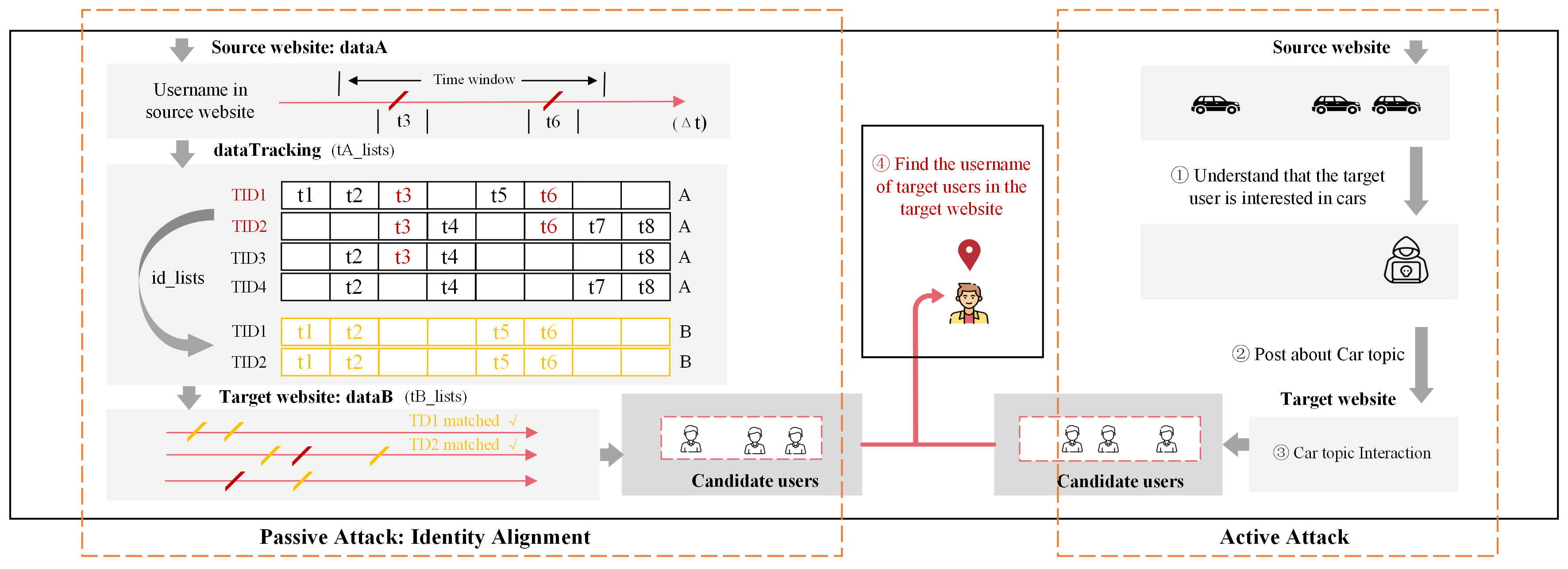}
  \label{fig:attacktwo}
  \vskip -0.15in
  \caption{Active and passive attacks} 
  \vskip -0.15in
\end{figure*}

(i) \textit{Passive attack} (straightforward, the \textit{left} side in Fig.\ref{fig:attacktwo}) aims to identify the potential username of a target user on a target website by analyzing publicly available web page data and tracker's anonymity data. Specifically, this method assumes that the target user C has accounts on both the source website A and the target website B. By analyzing the account AC on website A, it attempts to locate the corresponding account BC on website B, thereby identifying the user's identity on the target site.

(ii) \textit{Active attack} (advanced, \textit{right} in Fig.\ref{fig:attacktwo}) builds on the passive attack by taking the identified set of usernames on the target website and further investigating topics of interest to the target user. Relevant information is then posted on the target website to lure the user into frequently posting or interacting with the content. Subsequently, the users who engage with these topics are cross-referenced with the identified set of usernames, narrowing down the pool of potential target usernames even further.

\subsection{The Passive Attack}\label{sec-passiveattack}

The passive attack methodology harnesses our identity alignment algorithm and the generated anonymity dataset to effectively identify correlations between users across different platforms. By utilizing cookie synchronization mechanisms, tracker websites can monitor and record user behavior on both source and target websites. Our identity alignment algorithm, powered by tracker data, enables the de-anonymization of the target user while facilitating a detailed analysis of their account on the target website. This approach narrows down the list of potential users to a small, targeted set, enhancing the precision of the attack. 

We present the attack (short for \textbf{A}) as follows.

\begin{itemize}
    \item[A1] \textit{Initial setup}:
    On the source website, the user has an account with a known identity. The site collects detailed user activity data, including posts, likes, and retweets. On the target website, the user has an account with similar activities but remains anonymous.

    \item[A2] \textit{Data collection and tracking}:
        Trackers collect user activity data from the source site (known identity) and anonymized data from the website.
        Trackers deploy cookies and other online tracking technologies on both sites. Cookies track user behavior and interactions, and synchronization mechanisms link cookies between sites, enabling cross-site tracking.

    \item[A3] \textit{Behavioral data analysis and de-anonymization}:
         Attackers obtain the tracker's data through attack methods and use the identity alignment algorithm to analyze and match the behavior of the tracker with that of the website. As a result, the attacker is able to determine the target website username of the user.
\end{itemize}

\subsection{The Active Attack}\label{sec-activeattack}

In passive attacks, natural user behaviors are carefully observed and analyzed to narrow down the pool of potential target usernames. Active attacks then build on this by employing targeted induction techniques to further refine the list. The process involves intersecting the recent participants in relevant discussions on the target site with the previously identified usernames, thereby significantly narrowing the set of potential targets.

We present the attack (short for \textbf{A'}) as follows.

\begin{itemize}
    \item[A'1] \textit{Collecting user preference information from the source website as induced information}: Investigate topics of interest to users by passively identifying usernames on the target website.
    
    \item[A'2] \textit{Injecting induced information into the target website}: Once the target topics are identified, relevant content can be published on the target website. To encourage target users to leave comments,attacker can proactively @mention the target users under the content and pose questions, thereby prompting them to respond actively. Meanwhile, continuous follow-up is conducted on discussions involving target users—for example, further inquiring about details of target users' comments to prompt them to post content multiple times, thereby facilitating the generation of more user data.
    
    \item[A'3] \textit{Monitoring user access to induced information}: Closely monitor the dynamics of target users over an extended period and observe whether they are interested in the published topics. If active participation by target users is detected, include them in the priority screening list.
    
    \item[A'4] \textit{Narrowing down the range of suspicious user sets}: The Users who have recently participated in the topic on the target website can be cross-referenced with usernames identified through passive means. This further narrows down the user set, improving effectiveness.
\end{itemize}

\section{Evaluation}
\label{sec-metric}

\subsection{Evaluation Metrics (New)}\label{sec-algnmetri}

We consider two aspects: the success rate and the precision of the alignment, and formalize them into three derived metrics: the Identity Alignment Success Rate (IASR), Anonymity Set Scaling Rate (ASSR), and Accurately Identified User Propotion (AIUP).

\smallskip
\noindent\textbf{\ding{172} IASR: Identity alignment success rate.} IASR measures the probability that the algorithm correctly identifies a user's identity. It is an intuitive metric that reflects the algorithm's ability to avoid missed detections.
    \[
    \text{IASR} = \frac{\text{Number of successful alignment experiments}}{\text{Total number of experiments}}
     \]
    The success rate is primarily influenced by the time matching granularity: the identity alignment algorithm involves datasets from three websites (source SiteA, target SiteB, and tracking SiteT). Since timestamp synchronization can vary across different websites due to factors like tracking tag placement (tracking mechanism time difference) and time zones (host time difference), we use time matching granularity to address these discrepancies, aiming for fuzzy matching to mitigate the impact on alignment effectiveness.

\smallskip
\noindent\textbf{\ding{173} \textit{ASSR: Anonymity set scaling rate.}} ASSR (Anonymous Set Size Reduction) quantifies the effectiveness of the algorithm in reducing the size of the anonymous user set. It reflects the capability of the proposed alignment algorithm to minimize the search space, ensuring that the de-anonymized user set is as small as possible. The ASSR is calculated as follows:

\[
\text{ASSR} = \frac{\text{Size of the original anonymous set}}{\text{Average size of the de-anonymized set}} = \frac{|S|}{|S'|}
\]

\noindent where \( |S| \) represents the size of the original anonymous set, and \( |S'| \) is the average size of the de-anonymized set. The average size is determined by summing the sizes of the de-anonymized sets from each experiment and dividing by the total number of successful alignment experiments.

\smallskip
\noindent\textbf{\ding{174} AIUP: Accurately identified user proportion.} AIUP refers to the percentage of users whose de-anonymized set size is reduced to 1. This metric calculates the proportion of users within the dataset who are precisely identified. It focuses on the precision of the attack in correctly pinpointing individual identities without error. AIUP can reflect the effectiveness of a identity alignment algorithm in precisely determining user identities. Reducing a suspect's anonymous set to a single accurate target is crucial for the accuracy and efficiency of an investigation. AIUP provides a clear indication of the algorithm's capability to accurately narrow down the user set to a single identifiable entity, showcasing its effectiveness in identifying the identities of individuals involved in criminal activities.

\subsection{Experiment Configurations}

We utilized data from two prominent social networks for our experiments. The source website data consisted of user data from Twitter, crawled from December 17, 2007, to November 6, 2023, capturing usernames and posting timestamps. This dataset comprised over 10,000 Twitter users (10,655) and more than 40,000 posting times (46,546). The target website data encompassed user data from Sina Microblog, obtained from September 12, 2010, to November 6, 2023, through web crawling. The tracking data represents postings and browsing behaviors of users from both source and target sites, forming timestamp datasets. These datasets were generated using a data generation algorithm based on the source and target site data.

Given that the browsing-to-posting ratio and time difference ($\Delta t$) influence the effectiveness of the identity alignment algorithm, we modeled the number of views with a normal distribution, where the mean corresponds to the view-to-post ratio and the standard deviation is 1. Additionally, the power-law exponent of the browsing time offset is set to 2. This study's data generation process offers several advantages. Firstly, the data collection was conducted using publicly available web data, mitigating privacy concerns and providing benchmark evaluation data. Secondly, the chosen social networking websites represent a broad spectrum of platforms, encompassing both English-centric and Chinese-centric networks. This selection effectively simulates cross-border scenarios.

\subsection{Evaluation Results}\label{sec-experiment}

We validate the accuracy and effectiveness of our algorithms.  We mainly evaluate the impact of time matching granularity and user activity levels on metrics.

\subsubsection{Time Matching Granularity Impacts}
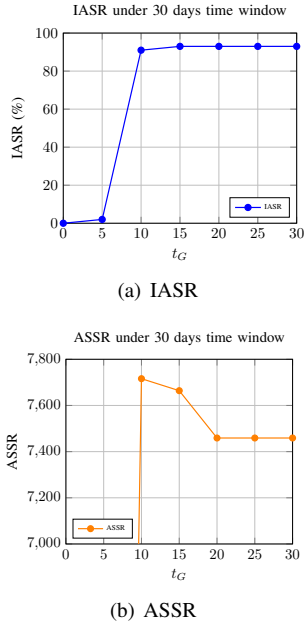
\begin{figure}[htbp]
    \centering
    \subfigure[IASR]{
\resizebox{0.47\linewidth}{!}{
    \begin{tikzpicture}
        \begin{axis}[
            width=7cm,
            height=6cm,
            xlabel={$t_G$},
            ylabel={IASR (\%)},
            ymin=0, ymax=100,
            xmin=0, xmax=30,
            xtick={0,5,10,15,20,25,30},
            ytick={0,20,40,60,80,100},
            grid=major,
            legend pos=south east,
            legend style={font=\tiny},
            title={IASR under 30 days time window},
        ]
        \addplot[blue, thick, mark=*] plot coordinates {
            (0, 0) (5, 2) (10, 91) (15, 93) (20, 93) (25, 93) (30, 93)
        };
        \legend{IASR}
        \end{axis}
    \end{tikzpicture}
    }
    }
    \subfigure[ASSR]{
   \resizebox{0.47\linewidth}{!}{
    \begin{tikzpicture}
        \begin{axis}[
            width=7cm,
            height=6cm,
            xlabel={$t_G$},
            ylabel={ASSR},
            ymin=7000, ymax=7800,
            xmin=0, xmax=30,
            xtick={0,5,10,15,20,25,30},
            ytick={7000,7200,7400,7600,7800},
            grid=major,
            legend pos=south west,
            legend style={font=\tiny},
            title={ASSR under 30 days time window},
        ]
        \addplot[orange, thick, mark=*] plot coordinates {
            (6, 0) (10, 7716) (15, 7664) (20, 7459) (25, 7459) (30, 7459)
        };
        \legend{ASSR}
        \end{axis}
    \end{tikzpicture}
    }
    }
    \vspace{-0.1in}
    \caption{IASR and ASSR under 30 days time window and different $\Delta t_G$ in data with browsing-to-posting ratio of 10 and $\Delta t$ of 10}\label{fig:granularity}
\end{figure}

In the course of a 30 days time window with a browsing-to-posting ratio of 10, we observed that (cf. Fig.\ref{fig:granularity}) as the time matching granularity increases, the anonymity set scaling rate decreases, while the alignment success rate increases. This is because a larger time matching granularity increases the probability of successful matches, but it also results in a larger pool of potential users, leading to a decrease in the anonymity set scaling rate.   

Furthermore, we observed a certain relationship between the time matching granularity and the dataset time difference. According to the experimental results as Table~\ref{tab:alignment-success-rate}, it can be seen that the alignment success rate only reaches a certain range when the time matching granularity is greater than the time difference in the dataset. This is because an increased time matching granularity, surpassing the time difference, is essential to alleviate biases caused by time differences and accurately identify users.

\begin{center}
\fbox{%
    \begin{minipage}{0.9\linewidth}

      \textbf{\text{Result-\ding{202}:}} With increasing time matching granularity, IASR increases while ASSR decreases. IASR reaches an acceptable rate when $\Delta t_G$ exceeds $\Delta t$.

    \end{minipage}
}
\end{center}

\subsubsection{User Activity Impacts.}
We conducted experiments to assess the impact of user activities, specifically posting density, under various real-world scenarios and parameters.

\smallskip
\noindent\textbf{Parameter selection.} We discovered that a critical factor affecting the scalability of anonymity sets is \textit{the density of user posting behaviors} within a given time window. Specifically, this density is determined by the number of user posts within an observation time window of a certain size. It has a great impact on the performance of our algorithm matching.

For instance, in the context of social networks like Twitter, user posting behavior reflects their level of activity. When many users post within a short time frame, the information density increases rapidly, making it challenging to accurately match user behavior patterns among a large number of users. However, directly selecting the users with the highest posting density may introduce outliers from users who post infrequently.

To address this issue, we focused on selecting the group of users with the highest posting density among those who are already highly active. We refined the time granularity to a single day and observed the maximum daily posting frequency for users (Table~\ref{tab:MaximumDaily}). By analyzing the data from 10,655 Twitter users, we found that 22 users had a maximum daily posting frequency of 20 posts. We considered these 22 users to be the most active.

\begin{table}[!htbp]
\caption{IASR obtained from different time granularity($\Delta t_G$) in different time difference($\Delta t$) datasets}
\vspace{-0.15in}
\resizebox{\linewidth}{!}{
\begin{tabular}{c|ccccccccc}
\diagbox{$\Delta t$ }{$\Delta t_G$ } & 0s & 5s & 10s & 15s & 20s & 25s & 30s \\
\midrule
0s & \cellcolor{gray!10} 97\% & \cellcolor{gray!10} 97\% & \cellcolor{gray!10} 97\% & \cellcolor{gray!10} 97\% & \cellcolor{gray!10} 97\% & \cellcolor{gray!10} 97\% & \cellcolor{gray!10}  97\% \\
10s & \cellcolor{gray!10} 0\% & \cellcolor{gray!10} 1\% & \cellcolor{gray!10} 85\% & \cellcolor{gray!10} 85\% & \cellcolor{gray!10} 85\% & \cellcolor{gray!10} 85\% & \cellcolor{gray!10}  85\% \\
20s &\cellcolor{gray!10}  0\% & \cellcolor{gray!10} 0\% & \cellcolor{gray!10} 0\% & \cellcolor{gray!10} 6\% & \cellcolor{gray!10} 86\% & \cellcolor{gray!10} 86\% & \cellcolor{gray!10}  87\% \\
30s & \cellcolor{gray!10}  0\% & \cellcolor{gray!10}  0\% & \cellcolor{gray!10}  0\% & \cellcolor{gray!10}  1\% & \cellcolor{gray!10}  3\% & \cellcolor{gray!10} 
 14\% & \cellcolor{gray!10}  76\% \\
\end{tabular}
}
\vskip -0.1in
\label{tab:alignment-success-rate}
\end{table}


\begin{table}[!htbp]
    \caption{Statistics of Maximum Daily Posting Quantity}
\vspace{-0.2in}
    \begin{center}
    \resizebox{\linewidth}{!}{
\begin{tabular}{c|ccccccc}

Max Daily Post  &  \cellcolor{gray!10} 20 &  \cellcolor{gray!10} 10 &  \cellcolor{gray!10} 5 &  \cellcolor{gray!10} 4 &  \cellcolor{gray!10} 3 &  \cellcolor{gray!10} 2 &  \cellcolor{gray!10} 1 \\
\cmidrule{1-5}
Number of Users &  \cellcolor{gray!10}22 &  \cellcolor{gray!10} 174 &  \cellcolor{gray!10} 161 &  \cellcolor{gray!10} 481 &  \cellcolor{gray!10} 1083 &  \cellcolor{gray!10} 2961 &  \cellcolor{gray!10} 5773 \\
    \end{tabular}
    }
    \end{center}
    \vspace{-0.1in}
\label{tab:MaximumDaily}
\end{table}


Furthermore, we recognize the critical importance of selecting the algorithm's \textit{time matching granularity} (as indicated in \textbf{Result-\ding{202}}) and the daily \textit{time window}.

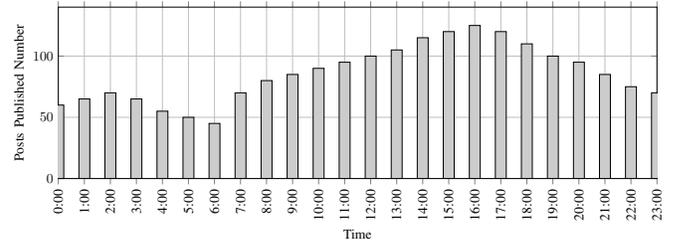
\begin{figure}[!htbp]
    \centering
\resizebox{\linewidth}{!}{
\begin{tikzpicture}
    \begin{axis}[
        ybar,
        width=17cm,
        height=6cm,
        xlabel={Time},
        ylabel={Posts Published Number},
        ymin=0, ymax=140,
        xmin=0, xmax=23,
        xtick={0,1,...,23},
        xticklabels={0:00, 1:00, 2:00, 3:00, 4:00, 5:00, 6:00, 7:00, 8:00, 9:00, 10:00, 11:00, 12:00, 13:00, 14:00, 15:00, 16:00, 17:00, 18:00, 19:00, 20:00, 21:00, 22:00, 23:00},
        bar width=8pt,
        grid=major,
        x tick label style={rotate=90, anchor=east}
    ]
    \addplot[
            ybar,
            color=black,
            fill=gray!40,
        ] coordinates {
        (0, 60) (1, 65) (2, 70) (3, 65) (4, 55) (5, 50) (6, 45) (7, 70) (8, 80) 
        (9, 85) (10, 90) (11, 95) (12, 100) (13, 105) (14, 115) (15, 120) (16, 125)
        (17, 120) (18, 110) (19, 100) (20, 95) (21, 85) (22, 75) (23, 70)
    };
    \end{axis}
\end{tikzpicture}
}
\vspace{-0.3in}
\caption{Distribution of posts published in 24 hours} \label{fig:24hour}
\end{figure}

To deepen our understanding, we conducted a statistical analysis of user data, as illustrated in Figure~\ref{fig:24hour}. This analysis examined the distribution of post counts within a 24-hour period based on our collected Twitter data. Using the day with the highest posting volume as an example (November 22, 2023), we observed that although the majority of posts were made between 9:00 and 1:00, the variation between the peak and lowest posting times was minimal. Consequently, we selected a time granularity of 60 seconds and a daily time window to thoroughly capture daily posting activities.

\smallskip
\noindent\textbf{Evaluation in two cases.} Once settle the parameters, another key challenge lies in the fact that, due to variations in tracker scripts across different websites, some trackers distinguish between browsing and posting behaviors when collecting mobile user data, while others do not. Therefore, we have analyzed and discussed these two cases separately.

\smallskip
\noindent{\textit{Case1: Distincting between browsing and posting behaviors.}}
\smallskip

We conducted experiments on users with different posting densities (20 posts/day, 10 posts/day, 5 posts/day, 4 posts/day, 3 posts/day, 2 posts/day, 1 post/day).  It is evident that (cf. \textit{upper} rows in  Table~\ref{tab:RUDSSASSRwithwithout}) users with lower posting densities have lower anonymity set scaling rates and lower precision in being accurately identified. However, even users with lower posting densities can still be accurately identified to some extent.

\begin{table}[!hbt]
    \caption{AIUP and ASSR at Different Posting Rates with and without distinction between browsing and Posting behavior}
    \begin{center}
    \vspace{-0.15in}
    \resizebox{\columnwidth}{!}{
    \begin{tabular}{c|c|ccccccc}

      \textbf{Method} &  \textbf{Posts/Day} &  \textbf{20} &  \textbf{10} &  \textbf{5} &  \textbf{4} &  \textbf{3} &  \textbf{2} &  \textbf{1} \\
\midrule
without & AIUP(\%) &  \cellcolor{gray!10} 78 &  \cellcolor{gray!10} 77 &  \cellcolor{gray!10} 72 &  \cellcolor{gray!10} 72 &  \cellcolor{gray!10} 59 &  \cellcolor{gray!10} 23 &  \cellcolor{gray!10} 20 \\
distinction & ASSR &  \cellcolor{gray!10} 8372 & \cellcolor{gray!10}  8071 & \cellcolor{gray!10}  7760 &  \cellcolor{gray!10} 7412 &  \cellcolor{gray!10} 6107 &  \cellcolor{gray!10} 3356 &  \cellcolor{gray!10} 3066 \\
\midrule
with & AIUP(\%) &  \cellcolor{gray!10}  91 &  \cellcolor{gray!10}  85 & \cellcolor{gray!10}  86 &  \cellcolor{gray!10}  79 &  \cellcolor{gray!10}  71 &  \cellcolor{gray!10}  65 &  \cellcolor{gray!10}  59 \\
distinction & ASSR &  \cellcolor{gray!10}  9016 &  \cellcolor{gray!10}  8954 &  \cellcolor{gray!10}  8972 &  \cellcolor{gray!10}  7923 &  \cellcolor{gray!10}  7679 &  \cellcolor{gray!10}  5429 &  \cellcolor{gray!10}  6314 \\

    \end{tabular}
    }
    \end{center}
    \vspace{-0.1in}
    \label{tab:RUDSSASSRwithwithout}
\end{table}

We extend our arguments based on these results. 

Considering that approximately 500 million tweets are sent daily on the Twitter platform, with free Twitter users restricted to a maximum of 20 posts per day, while some paid users can exceed this limit. Therefore, assuming an average of 20 posts per day per user, there are approximately 25,000,000 (twenty-five million) users posting each day. We calculate this based on a one-minute time granularity.

According to the Birthday Paradox, if we set the time granularity to 1 minute, with 24 hours in a day and 1440 minutes per day, for behavior trajectories involving 7 tweets per day, we would need around 150 million users per day to have a 99\% probability of finding two identical trajectories. This exceeds the daily posting user count. Therefore, users who post more than 7 times per day have a high probability of being accurately identified.

For users who post only once a day and can still be identified, our analysis revealed that among those accurately identified with one post per day, 46.8\% of them posted during the time period from 1:00 to 8:00, while users who could not be accurately identified had only 24.5\% of their posts during the same time frame. Therefore, it is attributed to the fact that some users who post only once strategically choose to share information during less active time periods, enabling them to be accurately identified.


\begin{center}
\fbox{%
    \begin{minipage}{0.9\linewidth}

      \textbf{\text{Result-\ding{203}:}} The accuracy of identity alignment improves with the increasing density of user posting behavior, particularly when using tracker data that distinguishes between browsing and posting activities

    \end{minipage}
}
\end{center}

\smallskip
\noindent\textit{{Case2: Non-distinction between browsing and posting behavior.}}
\smallskip

Some websites do not distinguish between posting and browsing behaviors in the data they send to trackers. Based on our investigation into the real-world data collected by these trackers, websites that differentiate between browsing and posting behaviors need to consider additional factors, such as the distribution of views per post (normal distribution with the mean as the browsing-to-posting ratio and an average difference of 1) and the distribution of browsing time offsets (power-law distribution with alpha=2).

\begin{figure}[!hbt]
  \centering
  \includegraphics[width=0.9\linewidth]{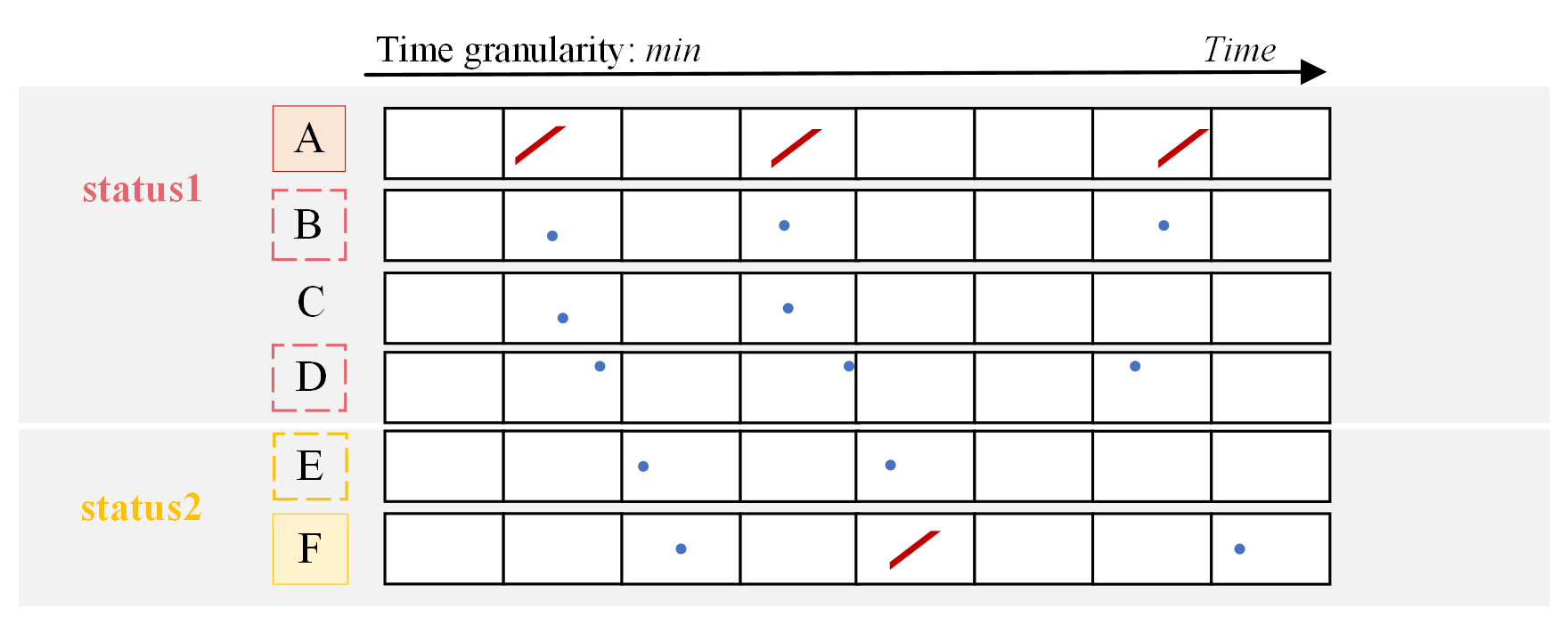} 
\caption{Status without Distinction (browsing vs posting behavior)} 
\label{fig:status12}
\end{figure}

When browsing and posting behaviors are not distinguished, a rise in the browsing-to-posting ratio leads to an increase in browsing activities that overshadow postings. Two scenarios can obscure posting records, as shown in Figure~\ref{fig:status12}: either the browsing record of the same post overshadows its posting record, or the browsing activities of other posts obscure the original posting record. This diminishes the scalability of the anonymity set and enlarges the de-anonymized set. Additionally, the discrete quantity of browsing activities also affects these rates.

We conducted identity alignment on users with varying activity levels, comparing the total number of users at each activity level with the number who could be precisely localized within that same level. Our experiments utilized a one-day time window and a time matching granularity of 60 seconds per user.

The experimental results (\textit{lower} rows in Table~\ref{tab:RUDSSASSRwithwithout}) indicate that users with lower activity levels exhibit smaller anonymity set scaling rates, and the proportion of users that can be precisely localized (with a de-anonymized set size of 1) is also lower. In contrast, users with higher activity levels are more likely to be precisely localized. However, even users with low activity levels (e.g., 1 post per day) have a certain probability (20\%) of being precisely localized.

\begin{center}
\fbox{%
    \begin{minipage}{0.9\linewidth}

            \textbf{\text{Result-\ding{204}:}} Using tracker data that does not distinguish browsing posting behavior, the accuracy of identity alignment improves as the density of user posting behavior increases.

    \end{minipage}
}
\end{center}

\subsection{Applying The Active Attack}

We conducted further experiments by applying an active attack on a target website with 10,655 users. 

Initially, a passive identity alignment attack was employed, yielding a candidate list of 10 usernames, resulting in an ASSR of 1066. To observe the effects of an active attack, we guided target users to interact with a topic of interest displayed on the source website. The following changes in ASSR, an evaluation metric for the accuracy of de-anonymization, were recorded:

\begin{itemize}

\item When 5 candidate users were led to send one message each, the ASSR increased from 1066 to 2131.
\item When 2 users were led to send two messages each, the ASSR rose to 5328.
\item When 1 user was led to send a single message, the ASSR reached 10,655.

\end{itemize}

Identity alignment is the process of identifying real identity behind anonymous data. We summarise three mainstream methods as follows. 

\begin{table*}[!htp]
    \centering
    \caption{Comparison of Related Work} \label{tab:relatedwork}
    \vspace{-0.15in}
        \resizebox{\linewidth}{!}{
        \begin{tabular}{c|p{5cm}|p{6.5cm}|p{5cm}}
       \toprule
            \textbf{\textit{Method}} & \textbf{Based on User Attributes} & 
 \textbf{Based on Network Structures} &   \textbf{Based on User Behavior}\\
        \midrule
            \textbf{\textit{Basics}} & Use self-disclosed information and statistical methods to align identities & Use algorithms and models to identify connections and similarities between users in different network structures. & Align identities by analyzing user behavior patterns and content generation\\
        \hline
            \textbf{\textit{Pros}} & Direct, simple, understandable & Accurate, large-scale, applicable to various network structures & Captures dynamic user behavior, suitable for behavior pattern analysis\\
        \hline
            \textbf{\textit{Cons}} & Low uniqueness of usernames, depends on user-disclosed personal information & High computational complexity, inefficient for large-scale networks, sensitive to changes in network structure & User may hide their behavior patterns, variations in content generation patterns affect accuracy \\
        \midrule
            \textbf{\textit{Studies}} & \cite{perito2011unique}\cite{liu2015user}\cite{li2019matching}\cite{lee2018second}\cite{motoyama2009seek}\cite{iofciu2011identifying}\cite{nagababu2024experimental}\cite{yin2024user} & \cite{singh2008global}\cite{brin1998pagerank}\cite{bayati2009algorithms}\cite{kuchaiev2010topological}\cite{gao2022dawn}\cite{zhang2016final}\cite{nassar2018low}\cite{xu2019gromov}\cite{heimann2018regal}\cite{du2019joint}\cite{chen2020cone}\cite{guan2025cross} & \cite{halimi2020profile}\cite{kusano2021user}\cite{ren2020banana}\cite{zhang2022behavior}\cite{li2018matching}\cite{srivastava2020words}\cite{nie2016identifying}\\
       \bottomrule
        \end{tabular}
        }
    \label{tab:related_work}
\end{table*}

\section{Related Work on Identity Alignment}
\label{sec-background}

\noindent\textbf{Method-I: User attributes.}
The method uses self-disclosed information, like usernames and locations, and employ statistical methods to determine if identities on different platforms belong to the same user.

Daniel et al.\cite{perito2011unique} calculated the probability of usernames belonging to the same user. Liu et al.\cite{liu2015user} proposed a weighted function based on various features of usernames to improve user identity recognition. Following this, Li et al.\cite{li2019matching} studied username patterns to identify users across multiple social networks.  Lee et al.\cite{lee2018second} used personal information like usernames and email IDs for identity alignment, specifically for personal homepage links. Motoyama et al.\cite{motoyama2009seek} proposed a matching method based on social relationship attributes, evaluating user match by calculating the overlap of attributes. Iofciu et al.\cite{iofciu2011identifying} proposed a user identity recognition method based on social tagging systems and user tagging behavior. Nagababu et al.\cite{nagababu2024experimental} proposed a cross-platform user identification method based on user attributes and behavioral history. Yin et al.\cite{yin2024user} proposed a method based on cross-attribute knowledge association, which embeds attribute associations in the knowledge graph and combines attribute recognition degree weighting calculation to improve the user entity alignment effect in sparse attribute scenarios.

User attribute methods provide accurate personal information but have limitations in fake profiles, low username uniqueness, limited information disclosure (e.g., criminals).

\smallskip
\noindent\textbf{Method-II: Network analysis.}
Network analysis use algorithms and models to identify connections and similarities between users in different network structures. 

The first approach is based on structural similarity (the homophily principle) to use personal relationships between users to achieve identity alignment across large-scale networks. 
IsoRank\cite{singh2008global} estimates node similarity by using an algorithm similar to PageRank\cite{brin1998pagerank} and pairs the two nodes with the highest similarity. NetAlign\cite{bayati2009algorithms} improve computational efficiency by formulating network alignment as a quadratic programming problem. However, GRAAL\cite{kuchaiev2010topological} is not scalable due to lengthy computational time. Korula and Lattanzi\cite{gao2022dawn} solve the alignment problem in social networks using local features, but only for specific networks.
When this method is applied to real social networks, many mismatched cases occur, reducing its practicality.
FINAL\cite{zhang2016final} improved alignment by introducing attribute information into the network structure. 
EigenAlign\cite{nassar2018low} employs a Low Rank approach to decrease memory usage. GWL technique matches graphs using node embedding and Gromov-Wasserstein\cite{xu2019gromov} difference but performs poorly on networks with significant structural variations.

The second approach leverages network embeddedness to represent network structure in real large-scale networks.
REGAL\cite{heimann2018regal} uses a similarity matrix of attributes and structural information to achieve network identity alignment through node embedding technology.
Du et al.\cite{du2019joint}  introduced a random walk algorithm that improves alignment accuracy through topological and structural similarity.
CONEAlign by Chen et al.\cite{chen2020cone} used DeepWalk and Graph Neural Networks to match nodes by learning node embedding vectors through random walks and GNNs. 
 Guan et al.\cite{guan2025cross} identified users across platforms by combining the user characteristics of social media platforms with the social network structure, and leveraging the similarity of their features and node embedding technology.
Static social network methods are effective, but can be limited by changes in network structure with dynamic social networks.

\smallskip
\noindent\textbf{Method-III: User behavior.}
User behavior and account information may show similarities in geographic activity and content. Location and time information are crucial for calculating user behavior similarity.

Liu et al.\cite{halimi2020profile} matched identities using login time and location, while Kusano et al.\cite{kusano2021user} established user associations by analyzing tweeting habits such as writing styles.
Ren et al.\cite{ren2020banana} developed a Temporal Graph Neural Network (TGNN) to predict user behavior across different social networks while addressing identity alignment with sparse anchor user datasets.
Zhang et al. achieved better results than similar studies by training a hierarchical memory network using user data from multiple social networks \cite{zhang2022behavior}.
Tracking criminals relies on users sharing their online behavior, but criminals hide their patterns. Behavior changes, reducing identification accuracy.
It is important to note that published content plays a significant role in aligning one's identity. Li et al.\cite{li2018matching} found similarities in user-generated content based on space, time, and writing style. They achieved an F1 value of 89.79\% by matching user behavior on social networks.
Srivastava et al.\cite{srivastava2020words} used tweet features like punctuation, emoticons, numbers, parts of speech, and high-frequency words to align identities. 
Nie et al.\cite{nie2016identifying} achieved identity alignment by analyzing users' interest tendencies on social networks, but this method also faces challenges.

Compared to existing solutions (cf. Table~\ref{tab:relatedwork}), our key innovation is the introduction of an active identity alignment method that enhances accuracy through user interaction, surpassing the capabilities of existing passive methods.

\section{Conclusion}

In this paper, we propose a novel identity alignment scheme to identify, measure, and analyze online hidden identities (e.g., criminal suspects). We implement and compare two de-anonymization attacks: a passive one based on tracking data and an active one that guides user interactions. We also establish a previously absent benchmark framework. Practically, we apply our scheme to a wide-ranging dataset and provide statistical results that mirror real-life scenarios.

\section{Deferred Evaluation for Datasets}
\label{sec-datasetEva}

As an add-on section, we further provide our evaluation methods and results (majorly regarding the accuracy) for generated anonymity datasets.

\subsection{How To Evaluate Our Dataset}

We first provide the way to evaluate our tracker's anonymity datasets (short for \textbf{D}).

\begin{itemize}

\item[D1] \textit{Perform web actions.} Engage in web activities, including browsing, logging in, posting, and interactions, and record the activities on the page.

\item[D2] \textit{Tracking data collection.} Visit the web page using Chrome collect data, and gather all request and  response data during the visit. Use Firefox to record function calls, property accesses, form events, and HTTP requests to third-party servers.

\item[D3] \textit{Tracking data annotation.} When third-party scripts are loaded on each page, analyze the requests and extract the target domains. EasyPrivacy is a filter list for online tracking that provides filtering rules to prevent trackers but does not offer specific tracker information. 
Use the tracker list from EasyPrivacy to verify whether the script or requested content contains trackers. Once it is determined that a tracking request domain is used, map the domain to the corresponding tracking platform according to the tracker info provided by Disconnect and Whotracks. 

\item[D4] \textit{Data screening \& classification.} Classify and screen the data according to JS scripts, cookie values, domain names, timestamps, and URL parameters to obtain valuable data. Both Disconnect and Whotracks provide specific information about trackers and advertisements on the website, including their names, types, and quantities. Specifically, mark sites with and without trackers. For sites with trackers, identify the tracker organization and locate the user's personal information in URL and cookies. 

\item[D5] \textit{Collecting public information on the web.} Gather information about user posting behavior on the web, including usernames and posting details such as timestamps, domain names, and other relevant information. Store the collected public information from different websites and pass it to the data analysis.

\item[D6] \textit{Data analysis.} Conduct statistical analysis based on detection indicators such as the presence of trackers, cookie synchronization, behavior types (e.g., browsing, posting), and the discrepancies between tracker records and public information.

\end{itemize}

\subsection{Evaluating Our Dataset}\label{sec-datasetEvaResults}

Then, we have collected 50 random sites, including 10 mainstream social networking sites. Randomly select 5-10 pages from the navigation bar of each site. Here, we highlight several key features of our collected data.

\smallskip
\noindent\textbf{\ding{172} Tracker and cookie synchronization.}
Through statistics, we found that 96\% (48/50) target and source websites have one or more trackers, among which 97.5\% (39/40) target websites have trackers, and 90\% (9/10) source websites have trackers. The number of pages with Cookie sync is 12\%(6/50). Trackers are commonly used to collect user behavior data on social networking sites like Twitter, Reddit, and Weibo. This data mostly consists of user browsing behavior, while posting dynamic behavior is relatively rare. Since trackers don't record all user behavior events, virtual identity alignment is affected.

\smallskip
\noindent\textbf{\ding{173} Discrepancy in recording types.}
Our empirical analysis covers social media platforms such as Twitter, Reddit, Weibo, and Douban, revealing the widespread usage of trackers to collect user browsing and posting behaviors. Due to functional differences among platforms, there is a variation in the ratio of user browsing to posting behaviors. Trackers may not comprehensively record all user activities.

\smallskip
\noindent\textbf{\ding{174} Time discrepancy between Web pages and trackers.}
Social networking sites (e.g., Twitter, Reddit, Weibo, Douban) use trackers to collect user behavior information such as browsing and posting dynamics. However, there are variations in the proportion of user browsing and posting behavior due to functional differences between websites. It's important to note that trackers may not record all user actions. Due to unsynchronized timestamps between websites and trackers, the time of user-posted content on a web page is denoted as $t1$, while the tracker records the time as $t2$. This time difference ($\Delta t$) varies across different scenarios, affecting identity alignment methods based on user-posted behaviors. We have identified four scenarios.

\begin{figure}[ht]
  \centering
  \includegraphics[width=0.99\linewidth]{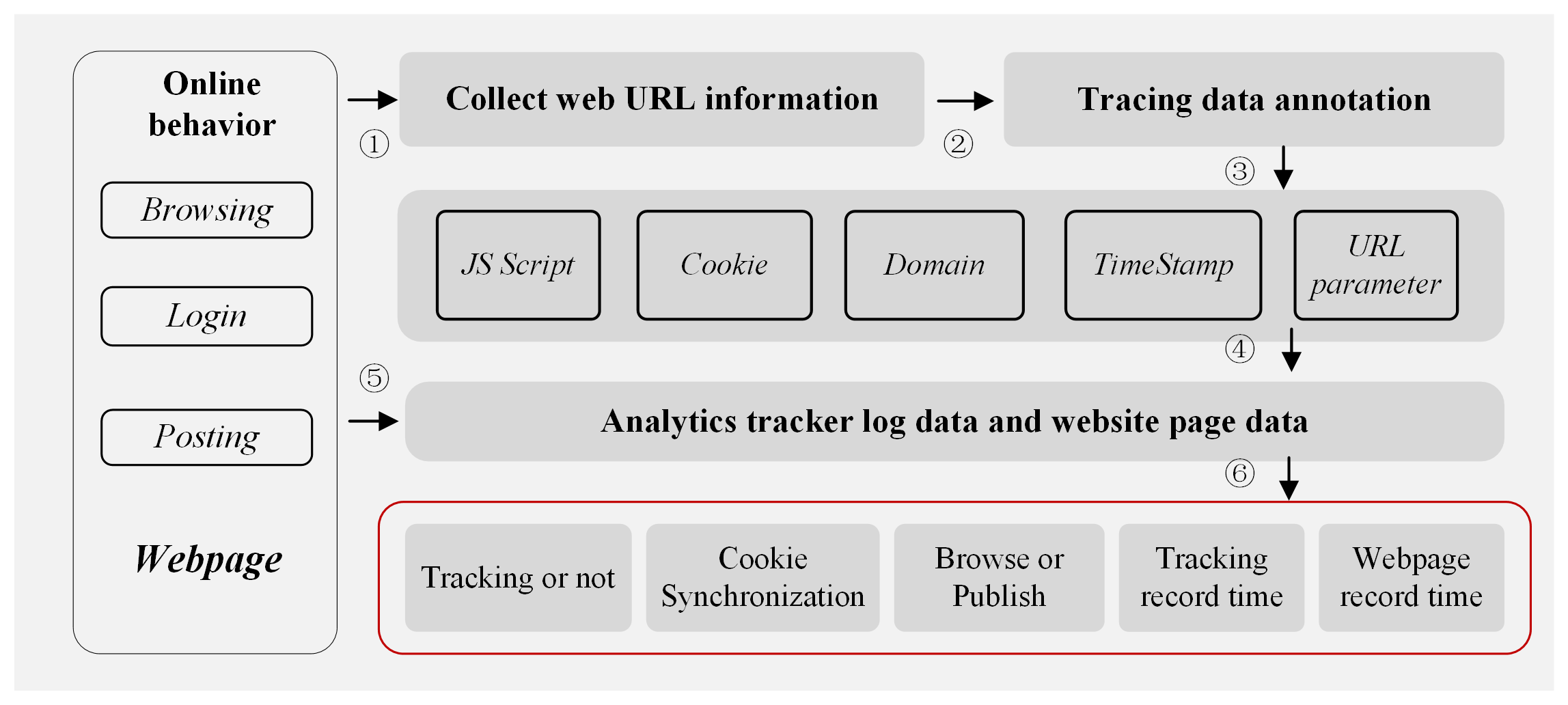}
  \vspace{-0.15in}
  \caption{Data Collection} \label{fig:dataanalysismethod}
\end{figure}

\begin{enumerate}
\item[$\bullet$] \textit{Case1.} The tracker accurately records the time and behavior of posting dynamic information when the user publishes a dynamic.  $\Delta t = t_2 - t_1 = 0$ 

\item[$\bullet$] \textit{Case2.} The tracker records the time when the user clicks the create dynamic button, not the time the dynamic is submitted.  $\Delta t = t_1 - t_2 =$ The time the user creates the dynamic, which is a dynamic variable. Therefore, the algorithm must consider the $\Delta t$ time variable. 

\item[$\bullet$] \textit{Case3.} The tracker records that the user has an activity but does not label it as a dynamic posting behavior.  $\Delta t = t_2 - t_1 = 0$. However, the tracker does not record the posting dynamic behavior, so the algorithm does not rely on behavior labels in this case. 

\item[$\bullet$] \textit{Case4.} When the user deletes a posted dynamic on the website, the tracker does not record it.  No $t_1$ exists. In this case, the time recorded by the tracker for posting dynamics can be considered as the browsing time. 
\end{enumerate}

\smallskip
\noindent\textbf{\ding{175} Distinguishing browsing and posting behaviors.}
The tracker Google on Reddit and the tracker Google on Douban do not distinguish between user behaviors, while the tracker Twitter on the Twitter website and the tracker Facebook Pixel on Facebook both distinguish between user behaviors\footnote{It is important to note that whether a tracker distinguishes user behavior is affected by the website on which it is deployed.}. For data that distinguishes between these behaviors, we generate browsing behavior data based on browsing quantity and browsing time offset.

 \begin{table}[htbp]
   \caption{With or Without distinction between browsing and Posting behavior}
   \label{tab:Posting behavior}
     \begin{tabular}{c|ccc}

     Website & Tracker & Without distinction & Distinction \\
     \midrule
     Twitter & Twitter &  \cellcolor{gray!10} &   \cellcolor{gray!10} $\checkmark$ \\
     Reddit & Google & $ \cellcolor{gray!10} \checkmark$ &  \cellcolor{gray!10} \\
     Facebook & Facebook Pixel & \cellcolor{gray!10} &  \cellcolor{gray!10} $\checkmark$ \\
     Douban & Google Analytics &  \cellcolor{gray!10} $\checkmark$ &  \cellcolor{gray!10} \\

     \end{tabular}
 \end{table}
    
\smallskip
\noindent\textbf{\ding{176} Distribution of views per post.}
User browsing behaviors towards different posts may exhibit natural randomness. We choose a normal distribution with the mean as the browsing-to-post ratio and a standard deviation of 1.

\smallskip
\noindent\textbf{\ding{177} Browsing time offset.} 
Browsing behaviors are more frequent shortly after post-publication. Browsing time offset follows a power-law distribution with an exponent of -2.

\section{Prototype: Identifying Crypto Suspects}
\label{sec-cryptoPrototype}

Illegal transactions and criminal activities involving cryptocurrencies are on the rise, posing significant challenges for law enforcement in tracing the identities of offenders. Our scheme can facilitate the implementation of a precise tracking system for blockchain. This system enables identity alignment, suspicious address verification, and the association of identity information with cryptocurrency addresses.

Our system consists of three key components. 
\begin{itemize}
    \item \textit{Suspicious crypto address verification}.
The system must allow users to verify whether a given cryptocurrency address is suspicious. It should help identify if the address is associated with criminal activities and provide detailed information about the specified cryptocurrency address.

\item \textit{Suspicious crypto user identity alignment}.
The system can perform identity alignment for users of suspicious cryptocurrency addresses. Users can choose source websites, target websites, and usernames, utilizing the system's embedded virtual identity alignment algorithms to obtain the results of the virtual identity alignment. This aids in understanding the set of suspicious users matched with the target website, which is beneficial for identifying those involved in criminal activities that cross the boundary between virtual and real identities. The process of virtual identity alignment should include the ability to align multiple external accounts, and information on users who have undergone virtual identity alignment should be recorded.

\item \textit{Linking identity information with crypto addresses}.
Finally, the system supports cross-checking specified cryptocurrency addresses, allowing users to provide a cryptocurrency address for verification against a suspicious address database. If the address is found to be involved in suspicious transactions, the system will output the associated identity information, assisting law enforcement in tracking and identifying individuals engaged in illegal activities.
\end{itemize}

We implemented the prototype using Bootstrap for the frontend and Python–Flask for the backend, with MySQL as the database to store address, tracking, webpage, and identity alignment data. The system is deployed on separate web and database servers, both configured with an AMD Ryzen 7 5800H (3.20 GHz) with Radeon Graphics. Experiments were conducted using a Chrome browser (v122.0.6261.70, 64-bit) in the same hardware environment.

We tested the prototyped system using a series of use cases that demonstrate the effectiveness of our scheme, i.e., successfully identifying the suspect's crypto identity by mapping a cryptocurrency address to their Weibo account. Our demonstration can be found at \textcolor{teal}{\url{https://github.com/lei20191116/AlignmentSys.git}}.

\section{Regulatory Compliance}
\label{sec-legal}

Although the data used in this method is sourced from public information, its cross-platform identity recognition feature still faces potential legal risks and ethical controversies due to varying legal frameworks across jurisdictions. This section therefore clarifies the scope of application and related disputes.

Ethically, cross-platform association of user identities via behavioral data may infringe on privacy, and the mechanism of actively inducing user interactions could violate the principle of "respecting user autonomy." Legally, unauthorized cross-border collection and use of data may contravene regulations such as the EU's General Data Protection Regulation and China's Personal Information Protection Law; additionally, the active attack module could be deemed illegal surveillance.

Thus, the application of this method is strictly limited to: use by judicially authorized law enforcement agencies (e.g., holding search warrants) for tracing criminal activities such as cryptocurrency money laundering and cross-border cyber fraud. It serves solely as a technical support tool for law enforcement and does not involve monitoring in civil or commercial domains. In practice, it must adhere to bilateral judicial authorization and platform collaboration policies to comply with legal requirements across different jurisdictions.

\bibliographystyle{IEEEtran}
\bibliography{bib}

\end{document}